# Crystallization kinetics and role of stress in Al induced layer exchange crystallization process of amorphous SiGe thin film on glass

Twisha Sain, Ch. Kishan Singh[*], S. Ilango and T. Mathews

Surface and Nanoscience Division, Material Science Group, Indira Gandhi Centre for Atomic Research, HBNI, Kalpakkam – 603102

[*] Electronic mail: kisn@igcar.gov.in; kisnsingh@gmail.com

| Authors | ORCID |
|---|---|
| Twisha Sain | 0000-0001-5989-2304 |
| Ch. Kishan Singh | 0000-0003-0243-6329 |
| S. Ilango | 0000-0002-5293-7802 |
| T. Mathews | 0000-0002-4876-8524 |

The present study reports Al induced crystallization (AIC) of amorphous (*a*)-SiGe in Al-Ge-Si ternary system at low temperature ~ 350 ºC. In addition to crystallization, the isothermal annealing of *a*-SiGe/AlO$_x$/Al/corning-glass (CG) structure was found to be accompanied by an Al induced layer exchange (ALILE) phenomenon. The evolution of residual stress in the Al layer during isothermal annealing is evaluated using X-ray diffraction based on modified $\sin^2\psi$ formalism to ascertain the role of stress in the ALILE process. A corroboration of the stress with the growth kinetics, analyzed using Avrami's theory of phase transformation gives a comprehensive understanding of the ALILE crystallization process in this system. The grown polycrystalline SiGe thin film is a potential candidate for novel technological applications in semiconductor devices.

# 1. INTRODUCTION

Over the years, continued efforts have been made to achieve polycrystalline (*poly*)-SiGe alloy thin film on glass and low cost substrates by bringing down the crystallization temperature ($T_{cryst}$). Such an achievement would enable significant cost reduction in photovoltaic industries [1, 2]. To this end, among many techniques, Al induced crystallization (AIC) process is an apt approach toward achieving this as AIC is successfully used to grow *poly*-Si and *poly*-Ge thin films on different substrates [3–5]. As a result, numerous studies on AIC of *poly*-SiGe have been widely reported over the years [6, 7]. Most of these studies employ a tri-layer structure with Al sandwiched between Si and Ge [8–10]. This approach involves a two-step process consisting of: - (i) diffusion assisted alloying of elemental Si and Ge, and (ii) subsequent formation of crystalline (*c*)-SiGe phase at temperatures ranging from 350 to 550 °C. In these studies, co-existence of either *poly*-Si or *poly*-Ge with the desired *poly*-SiGe thin film has been reported [9, 11–14]. Such phase segregation of *poly*-Si or *poly*-Ge phases can be largely impeded by using a pre-formed amorphous (*a*)-SiGe phase in the *a*-SiGe/Al bilayer structure [15, 16] and are extensively reported in literature [17, 18]. Yet some of the issues germane to crystallization kinetics of AIC in the ternary system (Al, Si and Ge) with *a*-SiGe/Al bilayer structure has not been explored adequately. This is particularly true for AIC process involving layer exchange and the origin of layer exchange during AIC process in this system has also not been investigated. Few existing reports attributes the cause of layer exchange phenomena that occurs during AIC in binary systems like *a*-Si/*c*-Al and *a*-Ge/*c*-Al to stress [19–21]. Some of the studies even attribute the low temperature crystallization process itself to stress [22]. The micro-structural changes that occur during ALILE crystallization process will inevitably affect and alter the state of residual stress of the overall thin film system, as it involves creation as well as re-arrangement of surfaces and interfaces between different phases. Hence, there is a need to ascertain the role that stress played in ALILE crystallization process. This motivate a systematic and quantitative investigation of residual stress that develop in the Al layer during AIC of *a*-SiGe/Al system. Subsequently, a correlation of the stress with the kinetics of crystallization will provide a comprehensive understanding of the ALILE crystallization process in *a*-SiGe/Al bilayer system.

In this report, we demonstrate the realization of *poly*-SiGe thin film on corning glass (CG) substrates at 350 ºC by employing AIC in *a*-SiGe/*poly*-Al bilayer. This $T_{cryst}$ is relatively lower than its bulk counterpart. Since the crystallization progresses with time under isothermal annealing conditions, Avarmi's theory of phase transformation is invoked to study the kinetics using Johnson-Mehl-Avrami-Kolmogorov (JMAK) equation [23]. The evolution of residual stress in Al layer during isothermal annealing is investigated using a method based on X-ray diffraction. The results are then discussed to develop a better insight into the origin of layer exchange that occurs during the crystallization process.

**2. EXPERIMENTAL**

For the present study, ~ 50 nm thick Al thin films were deposited onto cleaned CG substrate at room temperature (RT) using electron beam evaporation. The surfaces of the Al films were intentionally oxidized to form a diffusion controlling native $AlO_x$ layer (~ 1-2 nm thick) by exposing them to ambient for about an hour. Thereafter, ~ 100 nm thick *a*-SiGe layer was deposited onto these films using an e-beam evaporator to form *a*-SiGe/$AlO_x$/Al thin film architecture on CG substrates. The materials used for evaporation in the present study viz. Al (99.999 purity) and SiGe alloy (99.99 purity) were sourced from Testbourne Ltd (UK) and TAEWON SCIENTIFIC CO., LTD (Republic of Korea), respectively. The base pressure of the deposition chamber was ~ $10^{-8}$ mbar and the working pressure during deposition was maintained at ~ $10^{-7}$ mbar. The substrate stage was rotated during deposition to ensure film uniformity. The thickness ratio of *a*-SiGe/Al layer was maintained at 2:1 so as to achieve a conformal layer of poly-SiGe. The thicknesses of the individual layers were measured in-situ using a quartz micro-balance and the overall thickness of the film was verified by performing step height measurement with a stylus profilometer (DEKTAK). Post deposition isothermal annealing of the deposited samples were carried out at 350 ºC in $N_2$ ambience for up to 144 hrs duration ($t_a$). The heating rate was ~ 21 °C/min and the samples were cooled in flowing $N_2$ at ~ 1-3 °C/min to RT. The evolution in the nucleation and growth of the *c*-SiGe phases upon annealing were investigated using an optical microscope (Carl Zeiss, Axio Vert.A1) in both bright field as well as circular polarized light-differential interference contrast (C-DIC) modes. The phase characterizations of the pristine and annealed samples were carried out ex-situ using grazing incidence X- ray diffraction (GIXRD) and Raman spectroscopy. The GIXRD measurements were performed in a Bruker D8 Discover diffractometer with Cu Kα radiation (λ = 1.5406 Å) and 1.0° angle of incidence in parallel beam geometry at RT. The Raman measurements were performed in a Raman microscope (inVia, Renishaw, UK) using green light laser (514.5 nm) as the excitation source along with 1800 gr/mm grating and CCD detector in the backscattering configuration. The stoichiometry of the SiGe thin films were estimated using a field emission scanning electron microscopy (FESEM; Carl Zeiss, Supra 55) equipped with X-ray detector for energy dispersive spectroscopic studies (EDS).

## 3. RESULTS

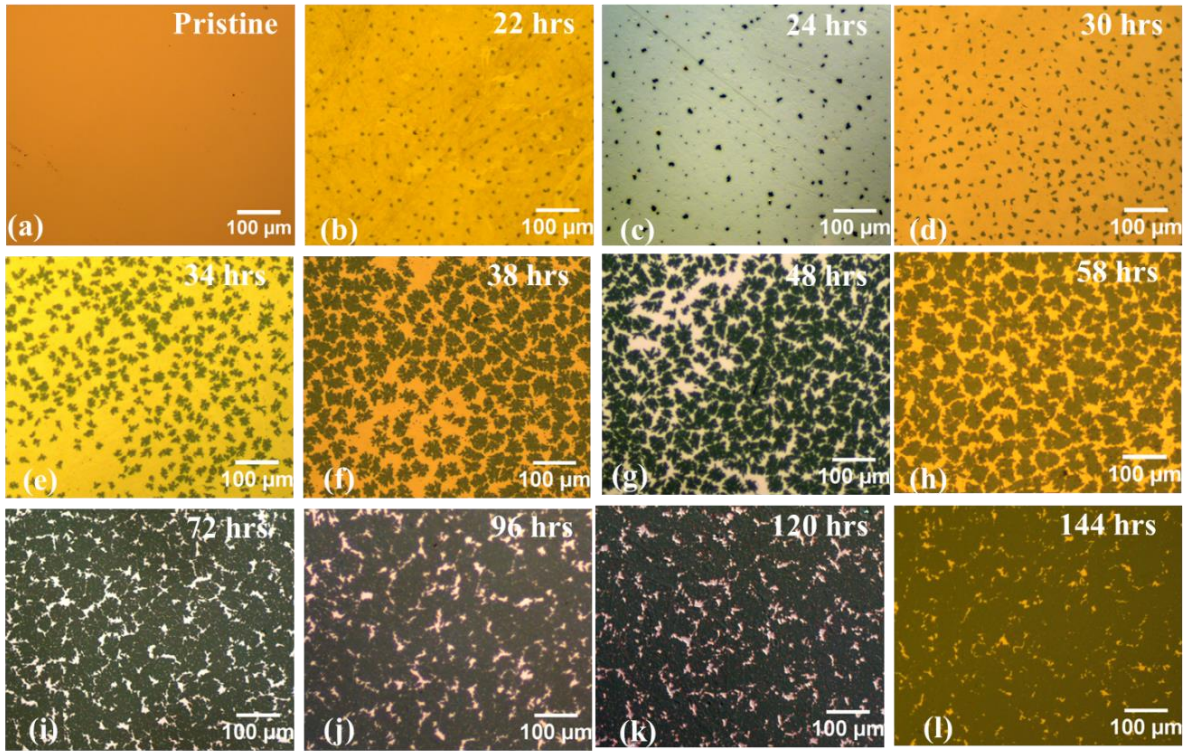

Fig.1 Optical micrographs of the *poly*-SiGe thin film on CG substrate(acquired from backside in C-DIC mode) after annealing at 350 °C for (a) Pristine; (b) 22 hrs; (c) 24 hrs; (d) 30 hrs; (e) 34 hrs; (f) 38 hrs; (g) 48 hrs; (h) 58 hrs; (i) 72 hrs; (j) 96 hrs; (k) 120 hrs and (l)144 hrs.

### 3.1 Results of annealing

Figure 1 shows the effect of isothermal annealing of the *a*-SiGe/Al/glass samples at 350 °C in $N_2$ ambient for various durations. The optical micrographs were acquired from backside of the glass substrates in C-DIC mode. The sequence of the images show the onset of ALILE and progress of Al catalyzed crystallization of the SiGe alloy thin film. The initiation of nucleation of the *c*-SiGe phase which are visible as nuclei/grains in the Al matrix starts after $t_a \sim 22$ hrs of annealing. The density of the nucleated *c*-SiGe grains in the annealed samples is plotted as a function of $t_a$ in Fig. 2. The plot shows that the nucleation of *c*-SiGe occurs abruptly at $t_a \sim 22$ hrs and the nucleation density remains almost constant with further increase in $t_a$. This clearly suggests that formation of the *c*-SiGe phase proceeds via a site saturated nucleation process. Subsequently upon increasing $t_a$, the *c*-SiGe grains continue to grow until they start to coalesce (above $t_a \sim 38$ hrs) to form continuous *poly*-SiGe thin film on glass. The growth kinetics of the ALILE crystallization process in these annealed samples is analyzed and discussed in later section.

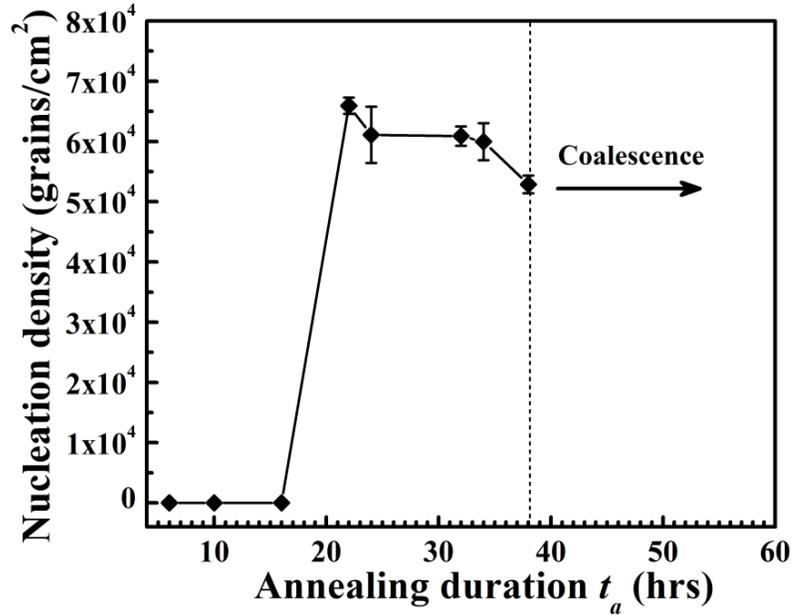

Fig.2 Nucleation density of the *c*-SiGe grains *Vs* annealing duration ($t_a$).

### 3.2 EDS results

The stoichiometry of the SiGe alloy thin film was measured using energy dispersive X-ray spectroscopy. The EDS spectra of the pristine, $t_a = 38$ hrs and $t_a = 144$ hrs annealed samples are shown in Fig 3. For $t_a = 38$ hrs sample, spectra acquired from both within the grain and grain free regions are shown in Fig. 3. The spectra are characterized by three main peaks at ~ 1.18, 1.49 and 1.74 KeV attributed to Ge $L\alpha$, Al $K\alpha$ and Si $K\alpha$ characteristic X-ray emissions (CXE). Because the overall thickness of SiGe/Al bilayer is only about 150 nm, the high energy (10 KeV) electrons are able to penetrate into the CG substrate and excite CXE from it. Hence, CXE from the CG substrate also appear in the spectrum as O $K\alpha$, Na $K\alpha$ and also contribute to Si $K\alpha$ peaks (Fig 3). Further, a measure of relative changes in integrated intensities of Ge, Al and Si peaks ($I_{Ge}$, $I_{Al}$ and $I_{Si}$) and their ratios (i.e., either $I_{Ge}/I_{Al}$ or $I_{Al}/I_{Si}$) upon annealing provides information on the changes in original layer structure of the pristine $a$-SiGe/Al bilayer [24]. The plot of relative change in $I_{Ge}$, $I_{Al}$ and their ratio $I_{Ge}/I_{Al}$ Vs $t_a$ is shown in Fig. 4. The $I_{Ge}/I_{Al}$ ratio in the pristine sample is ~ 3.01. This ratio is highest for the pristine sample as the Al $K\alpha$ line is heavily attenuated by the top SiGe layer (Al being the bottom layer). Any change in this layer structure due to diffusion of Ge (or Si) into Al layer or vice versa lead to decrease in $I_{Ge}$ (or $I_{Si}$) with a corresponding increase in $I_{Al}$, eventually, leading to the reduction of the $I_{Ge}/I_{Al}$ ratio. Now at $t_a$ ~ 38 hrs, there is a significant progress in the crystallization process (refer to Fig. 1). The $I_{Ge}/I_{Al}$ ratio reduces to ~ 2.4 for the data acquired from within the grains of $t_a$ ~ 38 hrs sample. This reduction indicates a change in the layer structure owing to initiation of layer exchange during crystallization of SiGe. Also note that the $I_{Ge}/I_{Al}$ ratio from the grain-free region (GFR) of the same sample is ~ 0.39. This is indicative of depletion of Ge (and Si) in these regions caused by diffusional migration of these atoms into the growing $c$-SiGe grains. The growth of $c$-SiGe grains in turn displaces and causes a reverse migration of Al to the top layer through the GFR and hence, $I_{Al}$ is quite high in these regions. Finally for $t_a$ = 144 hrs sample wherein the layer exchange is complete, the $I_{Ge}/I_{Al}$ reduces to ~ 1.18. The stoichiometry of the alloy thin film is estimated to be ~ $Si_{0.22}Ge_{0.78}$ after correcting for Si $K\alpha$ contribution from the substrate. The error in this estimation is expected to be ± 2%.

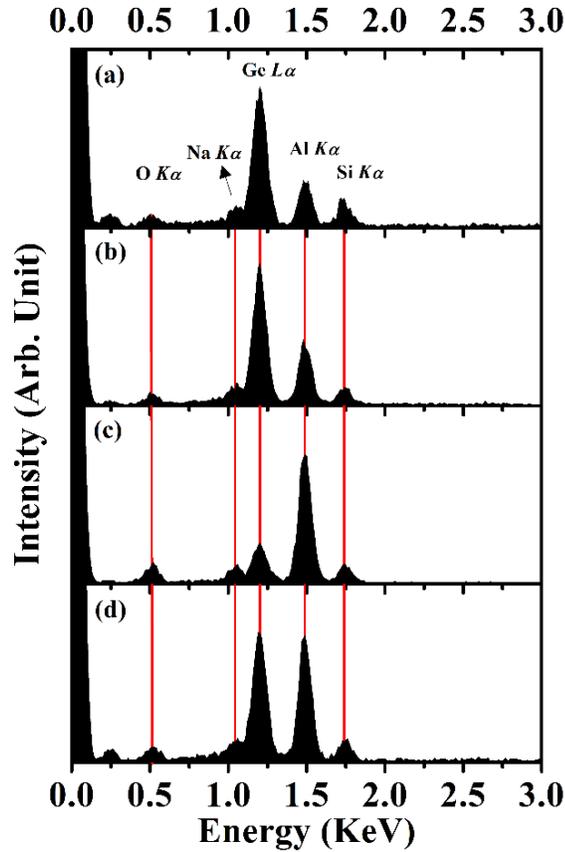

Fig. 3 (Color online) EDS spectra from: - (a) Pristine sample, (b) within grain of $t_a$= 38 hrs sample, (c) grain free region of $t_a$= 38 hrs sample and (d) 144 hrs annealed sample.

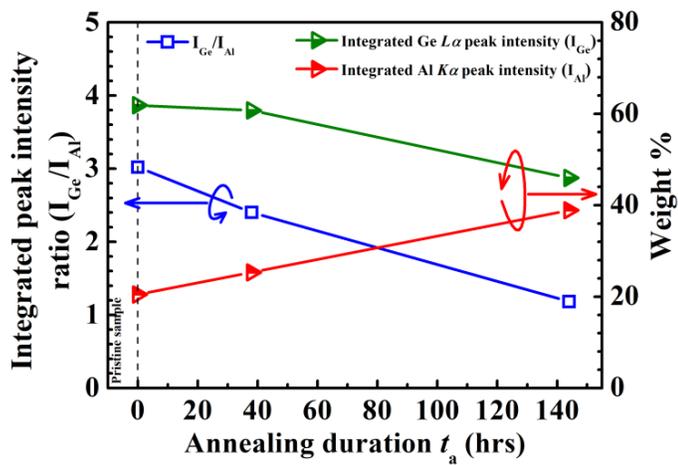

Fig. 4 (Color online) Plot of integrated intensities of the Ge and Al peaks ($I_{Ge}$ and $I_{Al}$) and their ratio $I_{Ge}/I_{Al}$ (estimated from EDS spectra of Fig. 3) *Vs* annealing duration ($t_a$).

## 3.3 Raman results

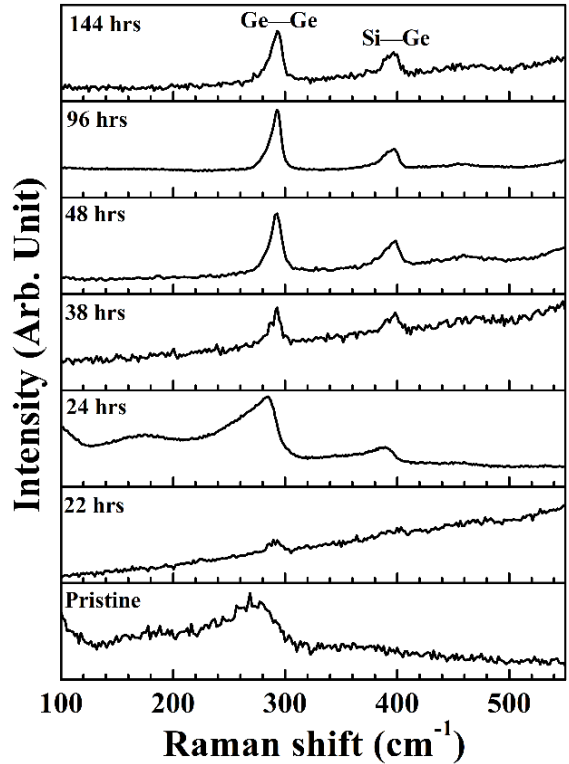

Fig. 5 Raman spectra of the pristine and the samples isothermally annealed at 350 °C (acquired from the backside) for various duration.

The CG glass substrates have very high percentage of transmittance (~ 92 %) for the wavelength of laser (514.5 nm) used in the micro-Raman spectroscopy. Micro-Raman also has high sensitivity over local microscopic regions. Hence, the phase structures of the $c$-SiGe nuclei/grains that grow in the initial stage of nucleation during the ALILE crystallization process can be effectively characterized by acquiring the Raman spectra from the backside of the sample through the glass substrates. Subsequently as the ALILE crystallization progress, $c$-SiGe phase grew to a volume detectable by GIXRD. The RT micro Raman spectra acquired from the pristine and annealed samples are shown in Fig. 5. The Raman spectra acquired from the pristine sample exhibit only broad peaks ~ 269 and 358 cm$^{-1}$ which are attributed to short range ordering among the Ge and Si atoms in $a$-SiGe matrix [25]. The new phase (nuclei i.e. seen as dark grain in Fig. 1) that nucleates after annealing the sample for $t_a$ = 22 hrs is confirmed to be $c$-SiGe phase from the Raman spectra. The spectra acquired within the nuclei or grain exhibit two peaks, one at ~ 292 cm [-1] and a broad peak at ~ 400 cm$^{-1}$. These peaks correspond to the vibration modes of Ge─Ge and Si─Ge bonds, respectively in *c*-SiGe [26–28]. Whereas the spectra obtained from the GFR show amorphous features similar to the pristine sample. The growth of the crystalline grains that is observed upon increasing $t_a$ ( Fig. 1) is reflected in the Raman spectra of annealed samples. The broad Si─Ge peak ~ 400 cm$^{-1}$ that is observed for $t_a$ = 22 hrs disappears and a well-defined peak corresponding to Si─Ge modes in *c*-SiGe appears at ~ 397 cm$^{-1}$ for $t_a$ = 38 hrs. Thereafter, the intensity of the Ge─Ge and Si─Ge modes increases with further increase in $t_a$ (beyond 38 hrs).The stoichiometry of the alloy thin film is estimated to be ~ $Si_{0.2}Ge_{0.8}$ from the Raman peak positions [27, 29].

### 3.4 GIXRD results

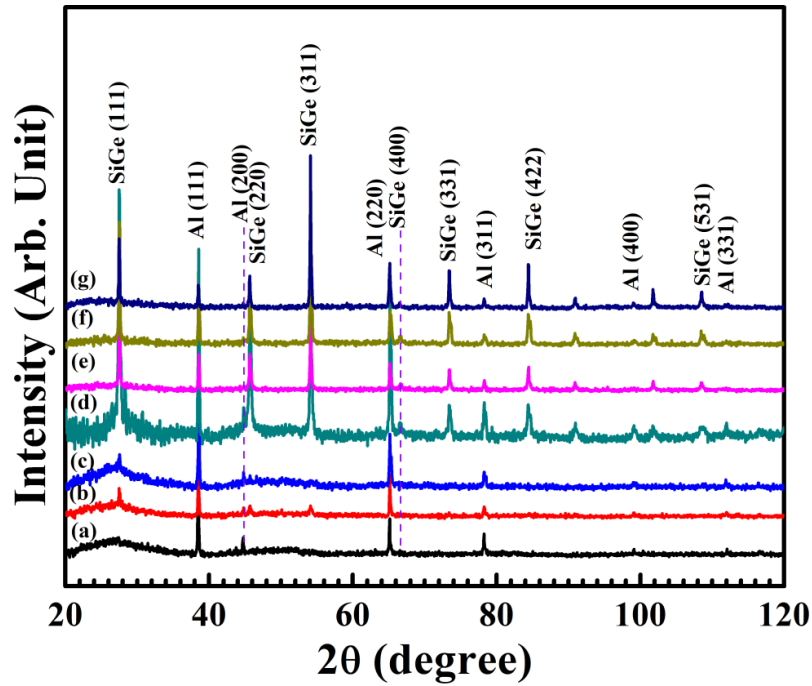

Fig. 6 (Color online) GIXRD pattern of (a) Pristine and the samples isothermally annealed at 350 °C in N$_2$ ambient for various durations viz., (b) 22 hrs; (c) 24 hrs; (d) 34 hrs; (e) 38 hrs; (f) 58 hrs and (g) 144 hrs.

The GIXRD patterns of the pristine and isothermally annealed samples (at 350 °C for various annealing durations) are shown in Fig. 6. The diffraction pattern for the pristine sample only exhibits peaks corresponding to *poly*-Al and no peaks pertains to *poly*-SiGe phase. This is expected since the pristine SiGe film is amorphous in nature. The peaks observed at 2θ ~ 38.48, 44.72, 65.12, 78.24 and 112.06 ° originate from (111), (200), (220), (311) and (331) planes of the *poly*-Al layer at the bottom. The GIXRD peaks corresponding to *poly*-SiGe phase can be observed only in annealed samples after the grains are visible (through optical microscope) subsequent to the onset of ALILE crystallization, which occurs in the sample annealed for $t_a$ = 22 hrs having an estimated crystallization fraction slightly above 3%. The crystallization is indicated by the small peaks observed at ~ 27.49, 45.67 and 54.09 ° those belong to (111), (220) and (311) planes of *poly*-SiGe, respectively (Fig. 6). The crystallinity of the *poly*-SiGe phase further improves with increase in $t_a$ as the remaining peaks pertaining to *poly*-SiGe start appearing prominently. This growth is confirmed by the peaks observed at 2θ ~ 27.48, 45.64, 54.10, 66.52, 73.42, 84.42, 90.91, 101.77 and 108.51° corresponding to the reflections from (111), (220), (311), (400), (331), (422), (511), (440) and (531) planes of *poly*-SiGe (Fig. 6).

## 4. DISCUSSIONS

The formation of *poly*-SiGe phase is clearly established using Raman spectroscopy and GIXRD. Now, *a*-SiGe/Al bilayer thin film forms a ternary system consisting of two semiconductors (Si and Ge) and Al. However, the ALILE crystallization process in *a*-SiGe/Al system still proceeds quite similar to those reported for binary Si-Al or Ge-Al system [30, 31]. There seems to be no heterogeneity in the way Si and Ge atoms diffuses into Al matrix after dissociative breakage from the *a*-SiGe phase. This is inferred from the direct nucleation of *c*-SiGe from *a*-SiGe phase without any formation of either *poly*-Ge or Si phase. The crystallization of *a*-SiGe phase into *poly*-SiGe phase at 350 °C is primarily due to the well-known phenomenon of metal induced crystallization (MIC) wherein Al acts as the metal[32]. The crystallization process is facilitated by the dissociative mechanism of the bonding network that prevails in $a$-Si$_{0.24}$Ge$_{0.76}$ that has excess of weaker Ge─Ge (1.9 eV) and Si─Ge covalent bonds compared to stronger Si─Si bonds (2.3 eV) [32]. The dominant process that mobilizes the Ge and Si atoms from *a*-SiGe phase to diffuse into Al is through dissociative breakages of the weaker Ge─Ge and Si─Ge covalent bonds. Consequently, the resulting *c*-SiGe phase has negligible directly bonded Si─Si bonds as can be seen from the almost non-existent localized Si─Si mode (normally observed between 450 to 500 cm$^{-1}$) in the Raman spectra (see Fig. 5)[33]. The crystallization kinetics begins with the nucleation of *c*-SiGe nuclei at the grain boundaries within the Al matrix. The grain boundaries in *poly*-Al serve as ideal sites for nucleation as both Si and Ge form eutectic with Al [34–37]. As the *c*-SiGe nuclei grow into bigger grains, measurement of the active area covered by *c*-SiGe grains give a direct estimate of the crystallized SiGe fraction ($f_{cryst}$). The evolution of $f_{cryst}$ in the annealed samples is plotted as a function of $t_a$ and it results in the S-shaped curve shown in Fig. 7. This curve contains information about the kinetic behavior of the crystallization process. The nucleation and growth phase are illustrated in Fig. 7. Since ALILE crystallization process is an isothermal solid-to-solid phase transformation, Avrami's theory of phase transformation can be applied to study its kinetics. The Avrami theory [38] relates the transformed crystalline phase to the annealing duration $t_a$ through the following Eqn.

$$f_{\text{cryst}} = 1 - \exp[-K(t^n)] \qquad (1)$$

where $t = t_a$ is the annealing duration, '$K$' is known as the Avrami constant and '$n$' is the Avrami exponent which determine the growth mode. Eqn. 1 is also known as Johnson-Mehl-Avrami-Kolmogorov (JMAK) equation.

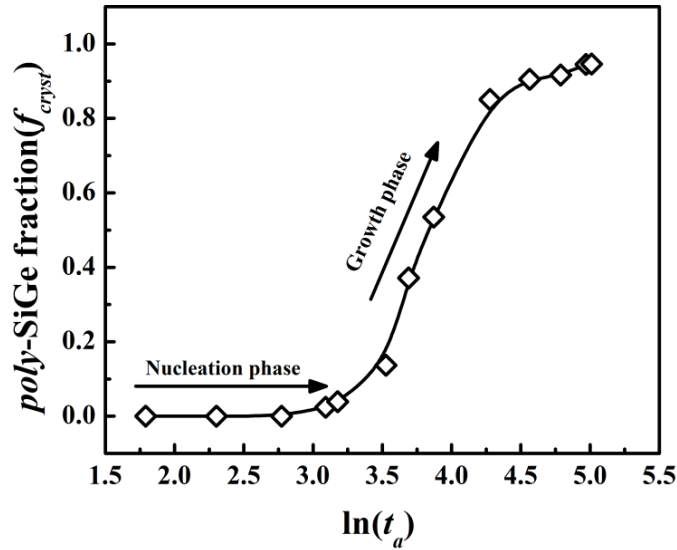

Fig. 7 Isothermal transformation plot - crystallized SiGe fraction *Vs* the logarithm of $t_a$ (the line is only a guide to eye)

To analyze the growth mode, the JMAK equation can be rearranged into

$$\ln[-\ln(1-f_{\text{cryst}})] = \ln K + n\ln(t_a) \qquad (2)$$

so that the slope of the $\ln[-\ln(1-f_{\text{cryst}})]$ *Vs* $\ln(t_a)$ gives the value of *n*. Also note that in a site saturated nucleation process (like in the present study), the value of *n* directly gives the dimensionality (D) of the growth [39]. The Avrami plot for the annealed sample is shown in Fig. 8. The plot is linear and exhibit a discontinuous jump between $t_a = 34$ and 38 hrs. As such, the Avrami plot exhibit two distinct linear regions marked as A and B. While the discontinuity arises because of an explosive crystallization that occur between $t_a = 34$ and 38 hrs, the change in slope (across the discontinuity) indicate a change in growth mode. For region-A i.e. the initial phase of growth just after nucleation, the Avrami exponent *n* is 3.04 (nearest integer = 3) suggesting a three dimensional (3D) growth in this region. Once nucleated, the *c*-SiGe nuclei (assuming to be spherical shape) follow a 3D growth until the diameter of the nuclei ~ approaches the thickness of Al film. Thereafter, the nuclei will cease to be spherical and assumes irregular shapes due to the dimensional constraint imposed by the architecture of the thin film. However, these nuclei (although irregularly shaped) continue to follow 3D growth as long as individual nuclei are well isolated. There is no significant growth in region-A. The maximum crystallization or $f_{\text{cryst}}$ (max) in this region is very less and limited to about ~ 12 % ($t_a$ ~ 34 hrs). In addition, the average distance of separation between the nuclei ($\lambda^{sep}$) is significantly larger than average size of the growing nuclei. For region-B, the Avrami exponent *n* is 1.87 (nearest integer = 2). The fractional value of *n* herein is attributed to the irregular shape of the dendritic grains. This shows that the growth kinetics makes a transition from 3D growth in region-A to 2D growth in region-B. During the transition, there is a significant grain growth and $f_{\text{cryst}}$ increase from ~ 12 % at $t_a = 34$ hrs to ~ 44 % at $t_a = 38$ hrs. As a result, $\lambda^{sep}$ reduces and becomes smaller than the average grain size. With further growth, the grains start coalescing in region-B. The coalescence of the growing grains reduces one growth dimension, curtailing an otherwise unhindered growth in region-B and make it a 2D growth. Coalescence leads to formation of interface and boundaries. Growth along these interfaces/boundaries is effectively 2D and much slower as well. Thereafter, the 2D growth continues till the crystallization is complete. Herein, it is pertinent to note that the process of ALILE crystallization also concurrently involves a layer exchange phenomenon between *a*-SiGe and *c*-Al layers. Hence, an comprehensive understanding of this crystallization process requires quantitative understanding of the mechanism that may lead to such layer exchange and we shall discuss it next.

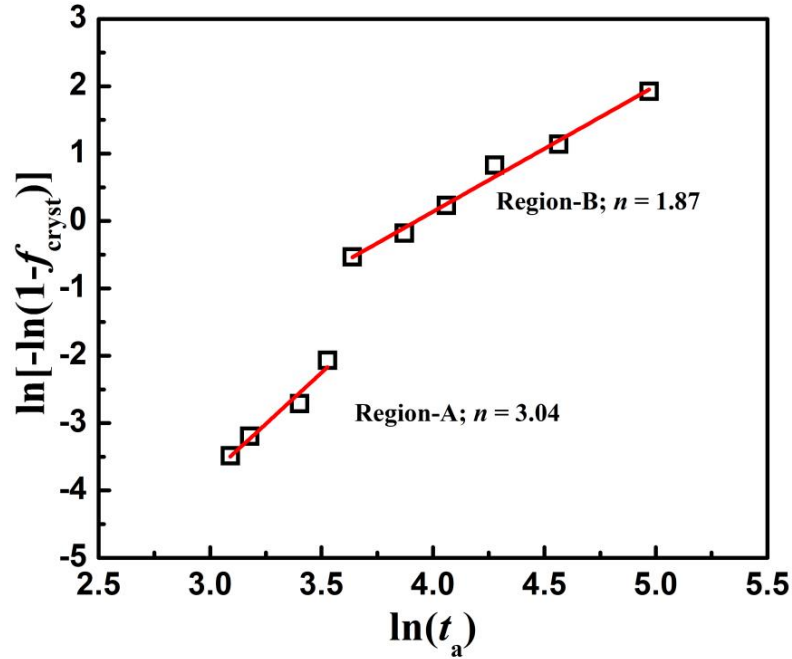

Fig.8 Avrami plot - ln [-ln (1-$f_{cryst}$)] $Vs$ ln ($t_a$)

In a solid-to-solid phase transformation (SSPT), strain energy inherently develops at the interphase boundary. This energy has to be taken into account while considering the free energy changes that drives the phase transformation(PT) and cannot be neglected like in case of a liquid-to-solid phase transformation (LSPT)[40]. Strain energy in a LSPT is generally neglected because the liquid phase can freely flow to accommodate any change in volume caused by the PT. The strain energy also can alter the state of residual stress in the transformed phase such that the system undergoes certain rearrangement during the PT or it can sometime drive the PT itself. To ascertain the role of stress in the layer exchange pertaining to the present study, evolution of stress within the Al layer during the AIC process needs to be investigated. For this purpose, in-plane residual stress $\sigma_\parallel$ in the Al layer is estimated from the GIXRD patterns (Fig. 6) using the multi $hkl$ $\sin^2\psi$ technique described elsewhere [41, 42]. In this approach, the sample tilt angle $\psi$ is defined as

$$\psi = \theta^{hkl} - \omega \tag{3}$$

where $\theta^{hkl}$ and '$\omega$' are the Bragg's angle for corresponding $hkl$ plane and the grazing incidence angle, respectively. The strain $\varepsilon$ is measured as

$$\varepsilon = \frac{(d^{hkl} - d^{hkl}{}_0)}{d^{hkl}{}_0} = \frac{(a - a_0)}{a_0} \qquad (4)$$

where '$d^{hkl}$' is the inter-planar spacing i.e., related to the lattice parameter '$a$' by the relation $a = d^{hkl}\sqrt{h^2 + k^2 + l^2}$ for cubic system. $a_o$ and $d^{hkl}{}_o$ are the strain free values of $a$ and $d$, respectively. $\sigma_\parallel$ (or $\sigma$) is then estimated from linear regression of the measured $\varepsilon_\psi$ w.r.t. $sin^2\psi$ according to the following relation

$$\varepsilon_\psi = \frac{1+\nu}{E}\sigma_\parallel sin^2\psi - \frac{2\nu}{E}\sigma_\parallel \qquad (5)$$

where $\nu$ and $E$ are the elastic properties viz., Poisson's ratio and Young's modulus, respectively. $\nu = 0.33$ and $E = 69$ GPa is used for Al in the present study for all calculations. The $a_o$ values were estimated from $a$ Vs $sin^2\psi$ plots using Eqn. 5 by setting $\varepsilon_\psi = 0$ [42]. The $\varepsilon$ Vs $sin^2\psi$ plot of the Al layer in the pristine and annealed samples are shown in Fig. 9.

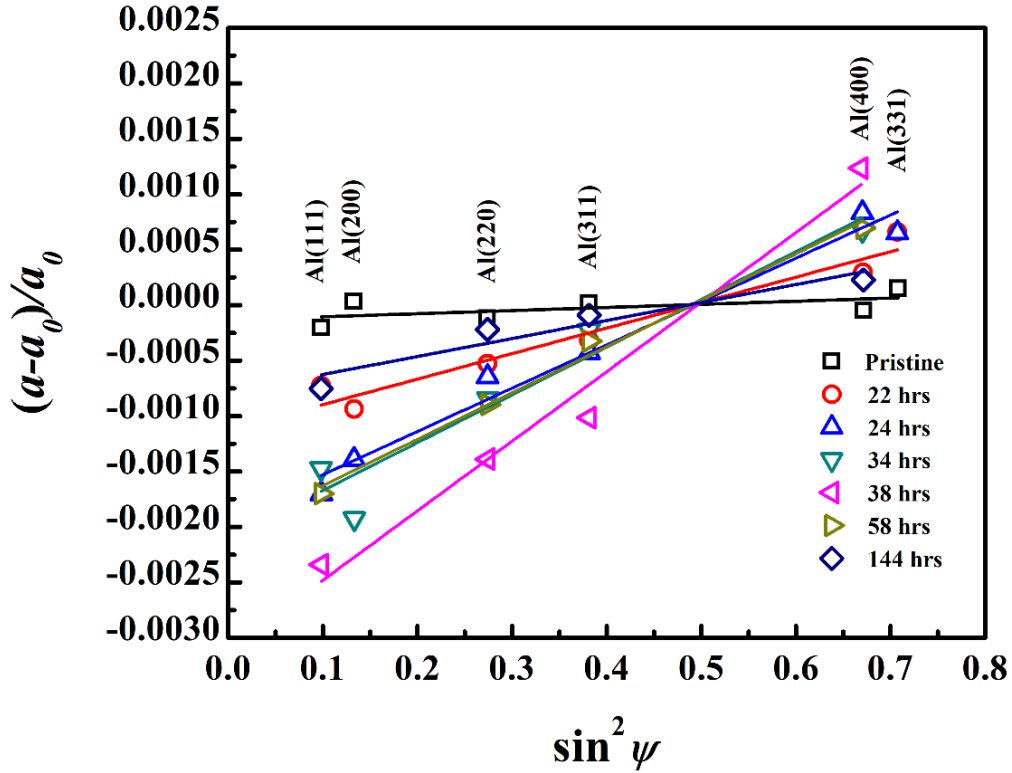

Fig. 9 Strain $\varepsilon_\psi$ Vs $sin^2\psi$ plots for the pristine and the samples annealed at 350 °C for various $t_a$.

Table 1: The $f_{\text{cryst}}$ and residual stress of pristine and the annealed samples. The error bar shown in the table is estimated from the standard error of the linear regression.

| SAMPLE | $f_{\text{cryst}}$(%) | RESIDUAL STRESS (MPa) |
|---|---|---|
| 1PRISTINE | 0 | 14.6 ± 9.8 |
| 22 hrs | 3.023 | 119 ± 13.3 |
| 24 hrs | 4.017 | 202 ± 15.5 |
| 34 hrs | 11.88 | 218 ± 27 |
| 38 hrs | 44.39 | 325 ± 32 |
| 58 hrs | 71.55 | 200 ± 26 |
| 144 hrs | 99.89 | 84 ± 18 |

In the $\varepsilon$ Vs $\sin^2\psi$ plot, a positive slope indicates an in-plane tensile stress in the thin film while negative slope indicate compression. The magnitude of stress calculated from the $\varepsilon$ Vs $\sin^2\psi$ plot are tabulated in Table 1. The Al layer in the pristine sample exhibits an in-plane tensile stress measuring about 14.6 ± 9.8 MPa. The origin of this tensile stress may be attributed to two main reasons. First the intrinsic growth stress that is inherently tensile at low thickness for most metal thin film [43, 44]. At ~50 nm (thickness), the film retains some of the tensile stress that built up during the grain boundary formation by coalescence. Secondly, out-of-plane compression in Al film due to deposition of *a*-SiGe layer leading to development of tensile component along the in-plane direction. Upon annealing the pristine sample, the tensile stress within the Al layer initially increases with increase in $t_a$ (see Fig. 10). This increase is due to the onset of diffusion of mobile Ge and Si atoms into Al layer during AIC process as well as the post growth of the Al grains [43–45]. It is accompanied by nucleation and concurrent growth of *poly*-SiGe phase within the Al layer (see Fig 1). The creation of this new crystalline interface between *c*-Al and *c*-SiGe increases the interface free energy. The interface energy ($\gamma$) has contribution from lattice mismatch (strain energy) as well as chemical interaction at the interface. During the nucleation and growth of *c*-SiGe in Al matrix, an Al GB is replaced by at least two or more *c*-Al/*c*-SiGe interfaces depending on where the *c*-SiGe nucleates. In terms of energy, this represents an increase from ~ 0.33 J/m$^2$ ($\gamma^{\text{GB}}_{\text{Al}}$ for Al GB) to ~ 0.76 J/m$^2$ ($\gamma^{\text{IF}}_{\text{Al/SiGe}} = 2 \times \gamma_{\text{Al/SiGe}}$ for *c*-Al/*c*-SiGe interfaces) [35, 46]. The residual stress that results from this interface strain is tensile in the Al lattice because of the larger *a* of SiGe compared to that of Al and higher average thermal expansion co-efficient (α) of Al compare to that of SiGe [47, 48]. As the $c$-SiGe fraction ($f_{cryst}$) increases within the Al matrix (via volume growth, note that grain density is constant), the stress contribution to γ increases because of the increase in total interface area. Thus, there is a corresponding increase in the magnitude of tensile σ in the Al layer with an increase in $f_{cryst}$. Subsequently as the stress builds up, the thin film system strives to lower $γ^{IF}_{Al/SiGe}$ and the overall γ. The system does so by driving Al to undergo a strain relaxing migration. Al gradually migrates away from the $c$-Al/$c$-SiGe interfaces to the voids left over by the diffusion of Si & Ge atoms (from the $a$-SiGe phase) to lower $γ^{IF}_{Al/SiGe}$. This leads to the initiation of layer exchange process. The evolution of tensile σ in the Al layer as a function of $f_{cryst}$ during the isothermal annealing is shown in Fig. 10. The ALILE is a gradual process and as such does not involve a well-defined critical threshold value of σ. During the initial 24 hrs of annealing, the tensile σ steadily rises from ~ 14.6 to 202 MPa while $f_{cryst}$ sees only a marginal rise from 0 to ~ 4 % (Fig.10). This shows Al layer could accommodate stress induced by the nucleation and subsequent growth of $c$-SiGe up to $f_{cryst}$ ~ 4%. In the next 10 hrs of annealing (i.e. $t_a$ = 34 hrs), the rate at which σ rises decline and increases nominally to ~ 218 MPa due to the initiation of strain relieving ALILE process. The $f_{cryst}$ ~ 12 % at $t_a$ = 34 hrs and 3D growth ensues up to this point as indicated by value of Avrami constant $n$ = 3 in this region. Then, the explosive crystallization that occurs between $t_a$ = 34 and 38 hrs cause a sudden rise in σ from 218 to ~ 325 MPa. The Avrami constant have anomalous values ($n > 5$) during the transition as shown in the plot of local Avrami exponent $Vs$ ln ($t_a$) in Fig 11. Thereafter, $n$ assume values ~ 2 and growth proceeds in 2D mode. Further, as $f_{cryst}$ increases with $t_a$ (> 34 hrs) in the 2D growth regime, Al layer relieves the strain caused by the additional increase in $f_{cryst}$ with the migrational mechanism explained above and σ starts decreasing. The ALILE process continues until the driving force for strain induced migration results in lowering of the overall γ. The ALILE crystallization also becomes increasingly slower in the 2D growth regime as $f_{cryst}$ approaches 100 % as the layer now has more *poly*-SiGe than the original Al film.

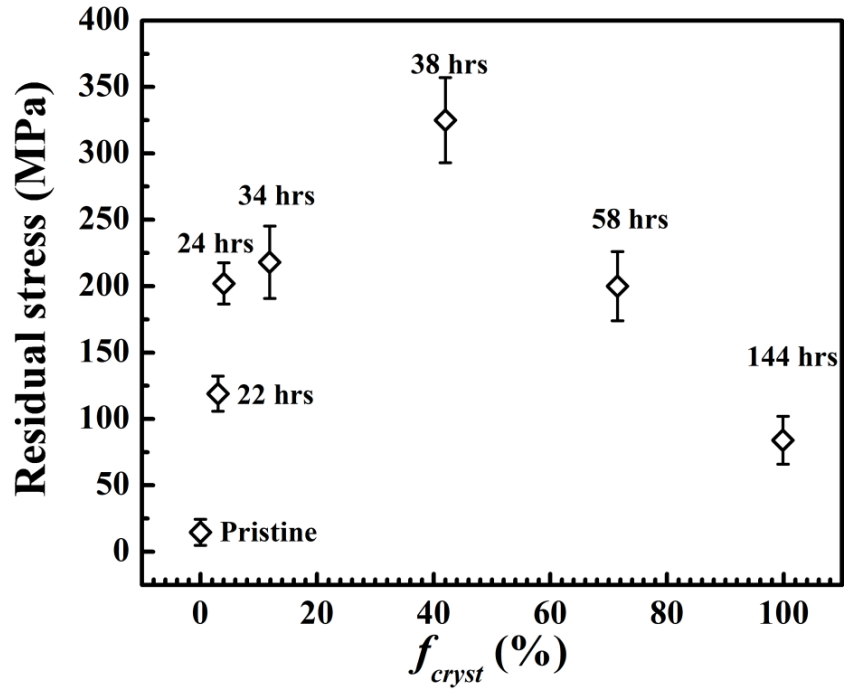

Fig. 10 Evolution of residual stress $\sigma_\parallel$ in Al layer w.r.t. the crystallized phase fraction ($f_{cryst}$) during ALILE crystallization process

Herein, we remind that all the XRD were performed at RT and that the stress estimated from it is residual stress. However, this residual stress can indirectly and reliably indicate the nature of stress that prevail during the isothermal annealing at 350 °C as explained in the following discussion. The intrinsic or growth part of the stress is ignored while considering stress evolution in the present thin film as they generally get relieved or become insignificant upon annealing at higher temperature [44]. This implies that both the in-situ stress that prevails during isothermal annealing at 350 °C and the residual stress measured at RT originate from thermal misfit stress that arises from the difference in α of the materials in contact [44]. It is strongly dependent on temperature and also on the rate of change of it. The $c$-Al/$c$-SiGe interface will be always accompanied with lattice misfit as and when this crystalline interface forms. This is because the value of α for Al (~ 21 to 25 ×$10^{-6}$ $K^{-1}$) is significantly higher than α for Si, Ge and SiGe (~ 2.9 to 6.9×$10^{-6}$ $K^{-1}$) for temperatures ranging from RT to 350 °C [47, 48]. The evolutionary trend of thermal stress during annealing and subsequent cooling in thin films can be predictably understood if α of the materials are known for the range of temperatures [49, 50]. So when the Al-SiGe system is heated at high rate ~ 21 °C/min, there will be large and rapid expansion of the Al lattice. The SiGe lattice at the *c*-Al/*c*-SiGe interface could only respond slower with a relatively lower thermal expansion leading to the generation of in-situ thermal stress that is compressive in the Al lattice. Subsequently upon cooling, tensile stress builds up in the Al lattice for similar reasons i.e., difference in α between Al and SiGe. The magnitude of this thermal tensile stress depends on the area of *c*- Al/*poly*-SiGe interfaces that dynamically changes as ALILE crystallization progress and also on the rate of cooling. While it is higher if the sample is cooled at high rate, a part of it will relax if cooled at lower rate. In the present study, residual tensile stress is measured at RT after cooling the sample naturally at ~ 1-3 °C/min. Similar studies on Al/Si and Al/Ge system shows that compressive in-situ stress builds up in the Al lattice during annealing and reverts to residual tensile stress upon cooling to RT [31, 51]. Hence, based on the observed tensile nature of the residual stress, the strain that actually drives the layer exchange process can be predicted to be compressive in the Al layer during the layer exchange process. The solid phase transformation of *a*-SiGe into *poly*-SiGe alters the state of stress in the Al layer. This strain driven layer exchange enable the growth of *poly*-SiGe on glass at 350 °C. The crystallization kinetics of the *poly*-SiGe shows that the initial growth phase starts with 3D mode and it switch over to 2D mode through an intermediate explosive growth. These films are good candidates for novel technological applications in solar and optoelectronic industries.

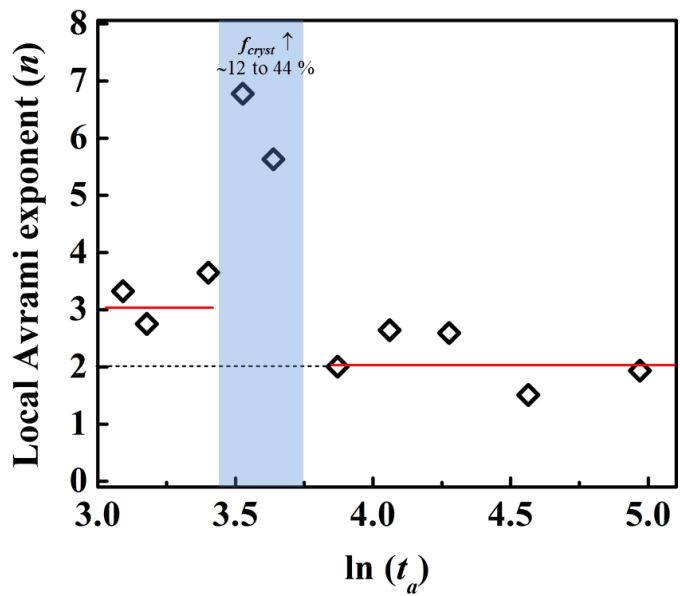

Fig. 11 Local Avrami exponent n *Vs* ln ($t_a$)

## 5. CONCLUSIONS

In conclusion, *poly*-SiGe thin films were achieved on glass at 350 °C using Al induced layer exchange (ALILE) crystallization process at temperatures well below the Al-Si & Al-Ge eutectics. The kinetics of the crystallization process was investigated and the role of residual stress in the layer exchange that occurs during the crystallization was also investigated. The driving force for the layer exchange process is the lowering of interface energies of the *c*-Al/*c*-SiGe interfaces that is created due to crystallization of *a*-SiGe into *poly*-SiGe in the Al matrix.

## ACKNOWLEDGMENTS

The authors would like to acknowledge Dr. S. R. Polaki, Dr. K. K. Madapu, Ms. Rajitha and Mrs. Sunitha for their help during the experiments and thank Dr. S. K. Dhara for careful reading and comments on the manuscript. The author also acknowledges UGC-CSR, Kalpakkam node for extending the facility for XRD and EDS measurements. The authors would also like to thanks Directors, MSG and IGCAR for their support in this work.

https://doi.org/10.1021/nl303801u